\documentclass[preprint,12pt]{elsarticle}

\usepackage{amssymb}

\journal{Physics Letters B}

\begin{document}

\begin{frontmatter}

\title{Critical Point on the QCD Deconfining Phase Boundary}


\author{P. K. Srivastava and C. P. Singh}

\address{Department of Physics, Banaras Hindu University, Varanasi 221005, INDIA}

\begin{abstract}
Ambiguities regarding the physics and the existence of the critical point (CP) on the QCD phase boundary still exist and the mist regarding the conjectured QCD phase boundary has not yet cleared. In this paper we extend our earlier study where we constructed a deconfining phase boundary using Gibbs' equilibrium conditions after using a quasiparticle equation of state (EOS) for quark gluon plasma (QGP) and an excluded volume EOS for the hadron gas (HG) and find the presence of a critical point on this phase boundary where the first order phase transition terminates. In this paper, we plot the difference in the normalized entropy density ($s/T^{3}$) between HG and QGP phases along the deconfining phase boundary and find that it vanishes at CP. Further we have shown the variation of the square of speed of sound ($c_{s}^{2}$) for the HG and QGP separately and find that the difference ($\Delta c_{s}^{2}$) between them along the deconfining phase boundary again vanishes at the CP of the boundary. We also plot the variation of shear viscosity to entropy density ratio ($\eta/s$) in HG as well as in QGP phases separately with respect to temperature at different baryonic chemical potential ($\mu_{B}$). The presence of a cusp like structure in $\eta/s$ again confirms the existence of CP on the deconfining phase boundary as predicted by some authors. These studies thus firmly indicate the precise location of CP as a second order phase transition point. 
\end{abstract}

\begin{keyword}

QCD critical point, Quasiparticle, QCD phase diagram
\end{keyword}

\end{frontmatter}



Precise mapping of the QCD phase boundary existing between hot, dense hadron gas (HG) and weakly interacting quark gluon plasma (QGP) and the location of a hypothesized QCD critical point (CP) have emerged as the most interesting and challenging problems before the experimental and theoretical heavy-ion physicists today [1-3]. The possible existence of CP in the temperature ($T$) and baryon chemical potential ($\mu_{B}$) plane of the QCD phase boundary was proposed a decade ago and it represents a second-order transition point where the first-order phase boundary terminates as $T$ further increases and $\mu_{B}$ decreases [3]. Its separation from the temperature axis $(\mu_{B}=0)$ spans the region of a cross-over transition where mesons in this densely populated region gradually break-up due to the thermal fluctuations and are finally converted into a gas of quarks, anti-quarks and gluons without any kind of phase transition. Lattice QCD studies employing Monte Carlo simulation, fail at $\mu_{B}\ne 0$ because the absence of a probability measure precludes direct computations and hence we often use mathematical approximations in order to determine QCD phase diagram for non-vanishing values of $\mu_{B}$. Some of these calculations have confirmed the existence of a cross-over chiral transition at $\mu_{B}= 0$ and $T= 170-200$ MeV and it ends at a CP with coordinates $\mu_{c}/3T_{c}\le 1.0$ [4-5]. However, other lattice calculations have cast a shadow of doubt on the existence of CP in the chiral phase transition [6]. Conclusively we still do not precisely know whether the conjectured phase boundary is an outcome of deconfinement and/or chiral symmetry restoration and whether the CP is related to the chiral phase transition alone [7]. Therefore, it seems interesting to study the nature of QCD phase transition and the existence of CP with the help of other phenomenological models. Recently we have constructed a first order deconfining phase boundary between HG and QGP phases employing Gibbs' criteria of thermodynamic equilibrium in a hybrid model approach [8-9]. These investigations further revealed an interesting finding that the deconfining phase boundary between HG and QGP terminates at a CP beyond which solutions of Gibbs' conditions are not possible indicating the presence of a cross-over transition. The precise location of CP in ($T$, $\mu_{B}$) plane was obtained as ($T_{c}=166$, $\mu_{c}=155$) when we used quasiparticle description for QGP [9] and this result is not much different from the location of CP obtained in the bag model EOS for QGP [8]. Our results get significant support from some earlier investigations where such hybrid model approaches [10-11] have been widely used. However, these studies do not reveal the precise nature e.g., the order of phase transition of CP. In this paper, we further extend our studies regarding the deconfining phase transition and the possible existence of CP by supportive evidences regarding the behaviour of the quantities such as normalized entropy density $s/T^{3}$, velocity of sound $c_{s}$, and the ratio of the viscosity to the entropy density $\eta/s$ near the CP in our hybrid model. In particular we calculate the difference in entropy density $\Delta s/T^{3}$ as well as difference in sound velocity $\Delta c_{s}^{2}$ between HG and QGP phases and we notice that these quantities vanish at the critical point and thus indicating a clear change in the order of the phase transition. By comparing the hadron-QGP transition with helium, nitrogen and water at and near their phase transition points in the $\eta/s$ ratio, Csernai, Kapusta and McLerran [12] have recently shown that the variation of $\eta/s$ with temperature in both the phases can pinpoint the location of the critical point. The studies of Sasaki and Redlich [13] and Lacey et. al. [14] lend further support to this finding that $\eta/s$ ratio reveals a cusp like feature when we study its variation with $T$ near the critical transition point between HG and QGP phases. Our aim in this paper is therefore, to explore further the precise location and the nature of CP on the deconfining phase boundary and to determine whether CP indeed is a point of second-order phase transition.\\

 Recently we proposed a new thermodynamically consistent, excluded-volume model for the hot and dense HG [8-9, 15]. Our approach incorporates the following new features. Besides thermodynamical consistency, our model uses full quantum statistics so that the phase boundary in the entire $(T, \mu_{B})$ plane can be investigated without using any additional approximation. Moreover, we incorporated excluded-volume correction arising due to a hard-core baryonic size only. We further assumed that the mesons can overlap and fuse into one another and hence cannot generate any hard-core repulsion. This is one big difference between our model and other models. We have demonstrated [15] that our calculations give a very good fit to the experimental ratios of multiplicities. Our model gives a proper description of all thermodynamical quantities including transport coefficients. The total pressure of the HG after incorporation of the excluded-volume effect can be written as [15-17]:
\begin{equation}
\it{p}_{HG}^{ex} = T(1-R)\sum_iI_i\lambda_i + \sum_j\it{p}_j^{meson}.
\end{equation}
In Eq. (1), the second term on the right hand side gives the total pressure from all mesons in the HG having clearly a pointlike size. $R=\sum_in_i^{ex}V_i^0$ gives the fractional occupied volume due to all types of baryons and $\lambda_i = exp(\frac{\mu_i}{T})$ is the fugacity of the particle. $I_{i}$ represents the integral involving the distribution function :
\begin{equation}
I_i=\frac{g_i}{6\pi^2 T}\int_0^\infty \frac{k^4 dk}{\sqrt{k^2+m_i^2}} \frac1{\left[exp(\frac{E_i}{T})+\lambda_i\right]},
\end{equation}
where $E_{i}$ is the energy ($E_{i}=\sqrt{k^2+m_i^2}$) with $k$ as the momentum and $m_{i}$ as the mass of the ith baryon. 

Our calculation for the shear viscosity is completely based on the method of Gorenstein et al. [18]. According to molecular kinetic theory, we can write the dependence of the shear viscosity as follows [19]:
\begin{equation}
\eta \propto n\;l\;\langle|{\bf k}|\rangle , 
\end{equation}
where $n$ is the particle density, $l$ is the mean free path, and hence the average thermal momentum of the baryons or antibaryons is:
\begin{equation}
\langle|{\bf k}|\rangle= \frac{\int_{0}^{\infty}k^{2}\;dk \;k \;{\bf A}}{\int_{0}^{\infty}k^{2}\;dk\;{\bf A}}, 
\end{equation}
and ${\bf A}$ is the Fermi-Dirac distribution function for baryons (anti-baryons). 
For the mixture of particle species with different masses and with the same hard-core radius $r$, the shear viscosity can be calculated by using equation [18]:
\begin{equation}
  \eta=\frac{5}{64 \sqrt{8} \;r^2}\sum_{i}\langle|{\bf k_{i}}|\rangle\times \frac{n_{i}}{n},
\end{equation}
where $n_{i}$ is the number density of the ith species of baryons (or anti-baryons) and $n$ is the total baryon density. 

To calculate the speed of sound at constant  $s/n$, we have used the recent formulation of Cleymans and Worku [20]. The speed of sound at zero chemical potential is easy to calculate where it is sufficient to keep the temperature constant [21, 22]. However, the speed of sound $(c_{s})$ at finite chemical potential can be obtained by using the following extended expression [20]:

\begin{equation}
c_{s}^{2}=\frac{\left(\frac{\partial p}{\partial T} \right)+ \left(\frac{\partial p} {\partial \mu_{B}} \right)\left(\frac{d\mu_{B}}{dT} \right)+\left(\frac{\partial p} {\partial \mu_{s}} \right)\left(\frac{d\mu_{s}}{dT} \right)}{\left(\frac{\partial \epsilon}{\partial T} \right)+ \left(\frac{\partial \epsilon} {\partial \mu_{B}} \right)\left(\frac{d\mu_{B}}{dT} \right)+\left(\frac{\partial \epsilon} {\partial \mu_{s}} \right)\left(\frac{d\mu_{s}}{dT} \right)},
\end{equation}
where the derivative $d\mu_{B}/dT$ and $d\mu_{s}/dT$ can be evaluated by using two conditions, firstly by keeping $s/n$ constant, and then imposing overall strangeness neutrality. Thus one gets [20]:
\begin{equation}
\frac{d\mu_{B}}{dT}=\frac{\left[n\left(\frac{\partial s}{\partial \mu_{s}}\right)-s\left(\frac{\partial n}{\partial \mu_{s}}\right) \right] \left[\frac{\partial L}{\partial T}-\frac{\partial R}{\partial T}\right] -\left[n\left(\frac{\partial s}{\partial T}\right)-s\left(\frac{\partial n}{\partial T}\right)\right]\left[\frac{\partial L}{\partial \mu_{s}}-\frac{\partial R}{\partial \mu_{s}}\right]} {\left[n\left(\frac{\partial s}{\partial \mu_{B}}\right)-s\left(\frac{\partial n}{\partial \mu_{B}}\right)\right]\left[\frac{\partial L}{\partial \mu_{s}}-\frac{\partial R}{\partial \mu_{s}}\right] -\left[n\left(\frac{\partial s}{\partial \mu_{s}}\right)-s\left(\frac{\partial n}{\partial \mu_{s}}\right)\right]\left[\frac{\partial L}{\partial \mu_{B}}-\frac{\partial R}{\partial \mu_{B}}\right]},
\end{equation}
and
\begin{equation}
\frac{d\mu_{s}}{dT}=\frac{\left[n\left(\frac{\partial s}{\partial T}\right)-s\left(\frac{\partial n}{\partial T}\right) \right] \left[\frac{\partial L}{\partial \mu_{B}}-\frac{\partial R}{\partial \mu_{B}}\right] -\left[n\left(\frac{\partial s}{\partial \mu_{B}}\right)-s\left(\frac{\partial n}{\partial \mu_{B}}\right)\right]\left[\frac{\partial L}{\partial T}-\frac{\partial R}{\partial T}\right]} {\left[n\left(\frac{\partial s}{\partial \mu_{B}}\right)-s\left(\frac{\partial n}{\partial \mu_{B}}\right)\right]\left[\frac{\partial L}{\partial \mu_{s}}-\frac{\partial R}{\partial \mu_{s}}\right] -\left[n\left(\frac{\partial s}{\partial \mu_{s}}\right)-s\left(\frac{\partial n}{\partial \mu_{s}}\right)\right]\left[\frac{\partial L}{\partial \mu_{B}}-\frac{\partial R}{\partial \mu_{B}}\right]},
\end{equation}
where $L=n_{s}^{B}+n_{s}^{M}$, is the sum of the strangeness density of baryons and mesons. Similarly $R=n_{s}^{\bar{B}}+n_{s}^{\bar{M}}$, stands for the sum of anti-strangeness density of baryons and mesons. In all the above calculations, we have taken an equal volume $V^{0}=\frac{4 \pi r^3}{3}$ for each baryon with a hard-core radius $r=0.8$ fm. We have taken all baryons and mesons and their resonances having masses upto $2 GeV/c^{2}$ in our calculation for the HG pressure. We have also used the condition of strangeness neutrality by putting $\sum_{i}S_{i}(n_{i}^{s}-\bar{n}_{i}^{s})=0$, where $S_{i}$ is the strangeness quantum number of the ith hadron, and $n_{i}^{s}(\bar{n}_{i}^{s})$ is the strange (anti-strange) hadron density, respectively [23]. \\

The EOS for QGP as used in this paper has been described in detail in Ref. [24, 25].  The detailed calcultaion regarding thermodynamical quantities like pressure, energy density, particle density etc. can be found in our earlier work [9]. We have demonstrated that our model agrees well with the detailed features of the curves obtained in the lattice QCD calculations. We are thus confident in using our QCD quasi-particle even at finite $\mu_{B}$ where lattice QCD fails miserably. We get a deconfining phase boundary after using Maxwell's construction and we find that the phase boundary as well as CP obtained by this method matches well with those obtained in Bag model calculation [8]. Here we give the prescription used by us for calculating the transport properties e.g., shear viscosity, speed of sound etc. Our calculation for shear viscosity is based on the prescription used by Sasaki and Redlich [26] where they calculate the shear as well as bulk viscosity for QGP in the quasiparticle model. The shear viscosity in a medium composed of one type of particle/antiparticle can be obtained from the following expression [26] :
\begin{eqnarray}
&&
\eta =  \frac{1}{15T}\int\frac{d^3k}{(2\pi)^3} 
\frac{k^4}{E^2}
\left[  g \tau f_0(1\pm f_0) 
{}+ \bar{g}\bar{\tau} \bar{f}_0(1\pm\bar{f}_0)  \right]\,, 
\end{eqnarray}
where $k$ is the momentum and $E=\sqrt{{ k}^2+M^2}$ is the energy with $M$ being the thermal mass for the quark and gluon. $\pm$ is for fermion and boson (i.e., quark and gluon), respectively and $f_{0}$ ($\bar{f}_{0}$) is the equilibrium distribution function for them, respectively and is given as :
\begin{eqnarray}
f_{0}(\bar{f}_{0}) = (e^{(E \mp \mu)/T}\pm1)^{-1}\,,
\end{eqnarray}
where we use $\mp \mu$ for $f_{0}$ and $\bar{f}_{0}$, and $\pm 1$ for quark and gluon, respectively. In Eq. (9), $\tau$ is the collision time which is determined by the thermal-averaged total scattering cross section $\langle v\sigma \rangle$, with the relative velocity of two colliding particles $v$ and the particle density $n$ in equilibrium. Further $g$ and $\bar{g}$ are the degeneracy factors for particle and antiparticle, respectively. We can get the shear viscosity for QGP simply by summing the Eq. (9) for all types of particles. The speed of sound for QGP is finally calculated by the following relation :
\begin{equation}
c_{s}^{2}=\left(\frac{\partial p}{\partial \epsilon}\right)_{V}=\left(\frac{\partial p/\partial T}{\partial \epsilon/\partial T}\right)_{V},
\end{equation}
where $p$ is the pressure and $\epsilon$ is the energy density of the QGP in our QPM.\\

In Fig. 1, we demonstrate the results obtained for the trace anomaly factor $\left(\epsilon-3p\right)/T^{4}$ in our hybrid model calculations using HG and QGP equation of state separately at $\mu_{B}=0$. We further compare our results with the results obtained in a recent lattice calculation [27]. We notice that our results yield an excellent fit to the lattice data. The success of our hybrid model which involves a separate and distinct description for both the phases (i.e., low temperature HG and large temperature QGP), is indeed excellent in reproducing the features of the lattice curve. Although such a combined description can still be treated as a crude one, but since lattice descriptions are still found quite unsatisfactory at $\mu_{B}\ne 0$, we emphasize that our hybrid model can provide an excellent phenomenological substitute for the formal theory.

\begin{figure}
\includegraphics[height=24em]{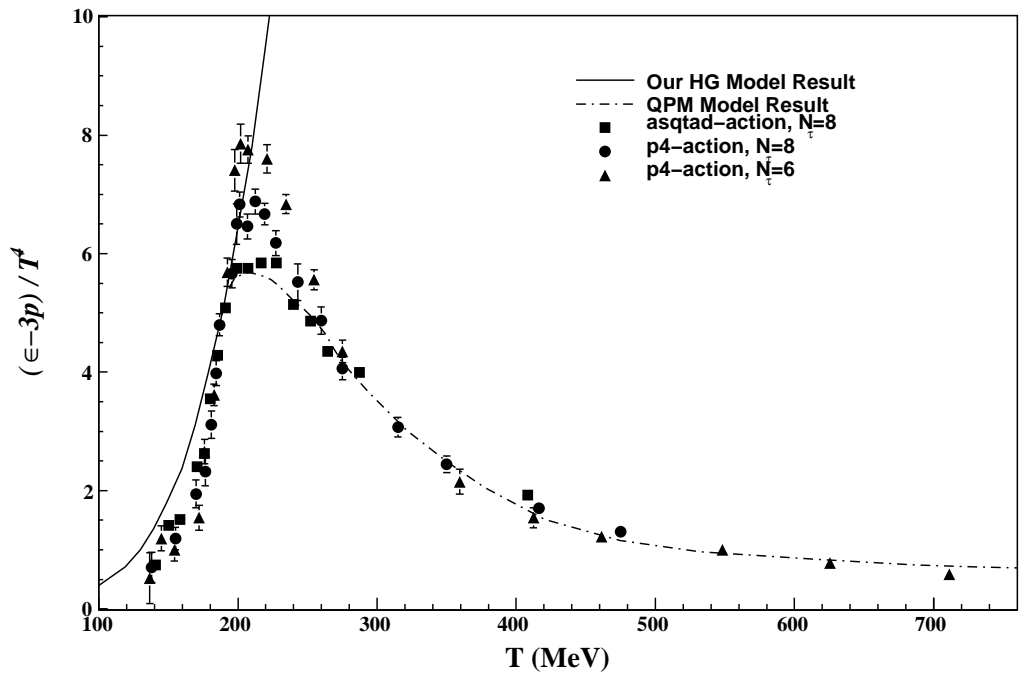}
\caption[]{Variation of trace anomaly $\left(\epsilon-3p\right)/T^{4}$ with respect to temperature. Solid line represents the low temperature behaviour of trace anomaly calculated in our excluded volume model for HG. Dash-dotted curve is the result obtained from quasiparticle model of QGP. Lattice data points are taken from Ref. [27].}
\end{figure}

\begin{figure}
\includegraphics[height=28em]{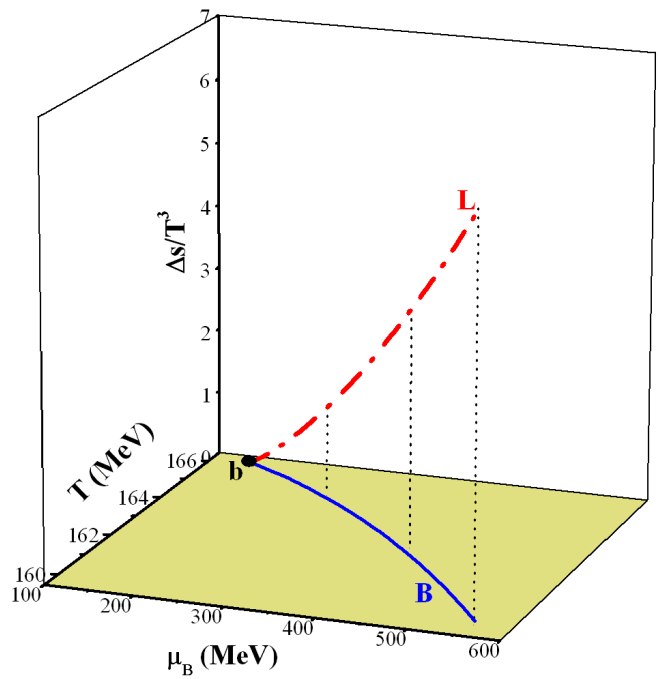}
\caption[]{Variation of $ (\Delta s/T^{3})=(s/T^{3})_{QGP}- (s/T^{3})_{HG}$ with respect to coordinates of various phase transition points on the $(T, \mu_{B})$ phase boundary. We have used transition points from Ref. [9].}
\end{figure}

\begin{figure}
\includegraphics[height=28em]{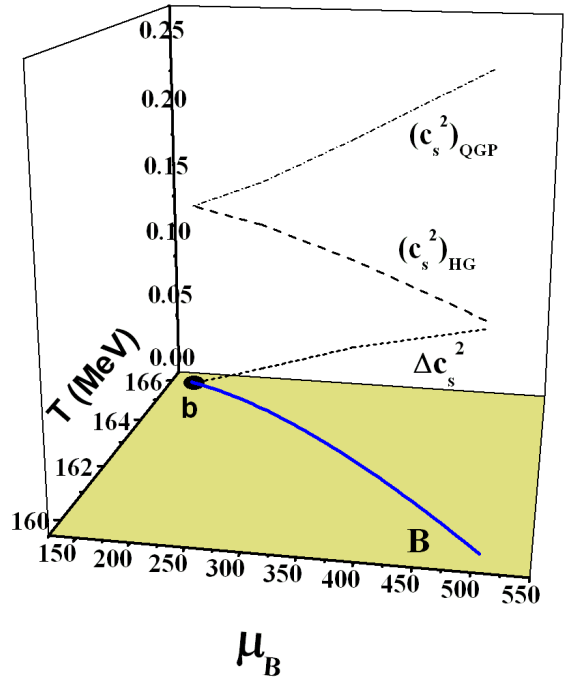}
\caption[]{Variation of $c_{s}^{2})_{QGP}$ (dash-dotted line ), $(c_{s}^{2})_{HG}$ (dashed line), and $ (\Delta c_{s}^{2})=(c_{s}^{2})_{QGP}- (c_{s}^{2})_{HG}$ (short-dashed line) with respect to coordinates of various phase transition points on the $(T, \mu_{B})$ phase boundary. We have used transition points from Ref. [9].}
\end{figure}

\begin{figure}
\begin{center}
\includegraphics[height=24em]{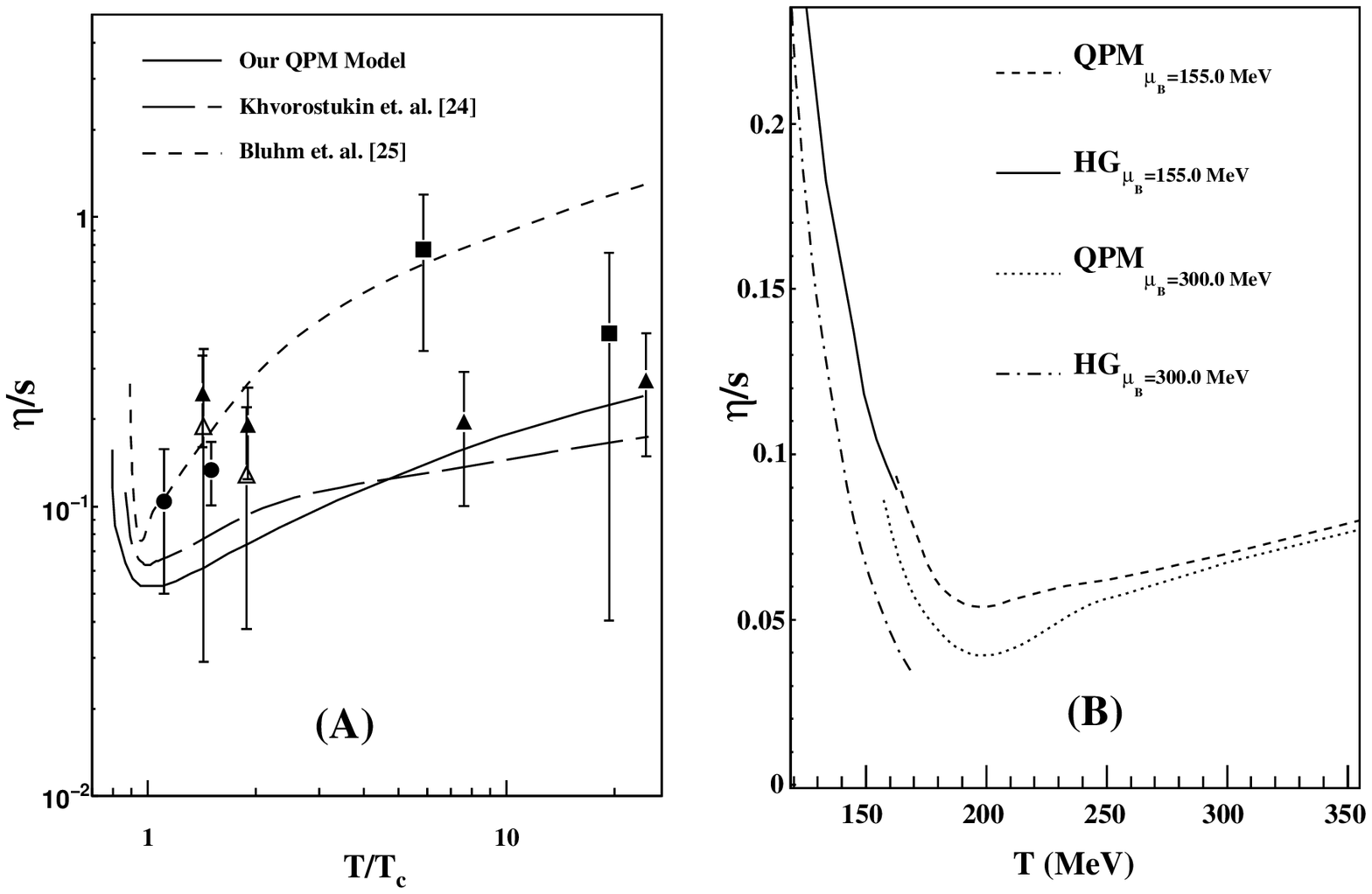}
\caption[]{(A) Variations of shear viscosity to entropy density ratio ($\eta/s$) for gluon plasma with respect to $T/T_{c}$. Solid line is the result from our calculation. Long-dashed line is the result obtained in Ref. [35] and short-dashed line is taken from Ref. [36]. The lattice data with $16^{3}\times 8$ and $24^{3}\times 8$ lattice are from Refs. [37] (triangles and squares) and [38] (filled circles). (B) Variations of $\eta/s$ obtained in our calculations for HG and QGP with respect to temperature at various $\mu_{B}$.}
\end{center}
\end{figure}

In Fig. 2, we have attempted to show what happens to the change in the entropy density at the CP of the phase diagram. We have calculated the difference $\frac{\Delta s}{T^{3}}=(s/T^{3})_{QGP}-(s/T^{3})_{HG}$ and demonstrated its variation with respect to the coordinates of the phase transition points lying at the deconfining phase boundary. We find that $\frac{\Delta s}{T^{3}}\ne 0.0$ and positive along the deconfining phase boundary in the case of first order phase transition which further indicates the presence of nonvanishing latent heat involved in the phase transition from HG to QGP. However, we surprisingly notice that $\frac{\Delta s}{T^{3}}\approx 0$ at the CP and thus the CP can be taken as a point where the first order phase boundary terminates and phase transition changes its order. This is certainly an interesting finding. Although we have used a hybrid model description for QGP and HG phases, the vanishing of net entropy density at the CP indicates that there exists a continuity in these two model descriptions. This means that the parameters in both descriptions had been suitably adjusted. 

Sound velocity is an important transport property of the matter created in nucleus-nucleus collision experiemnts because the hydrodynamic evolution of this matter strongly dependes on it. Speed of sound is related to the speed of small perturbations produced in its local rest frame. A minimum in the $c_{s}$ has also been interpreted in terms of a phase transition [20-21, 28-33] a large number of degrees of freedom is present which causes difficulty in the propogation of the sound wave in the medium. Further, Chojnacki and Florkowski [34] proposed that a shallow minimum in the speed of sound near the smooth joint of HG and QGP phases in a hybrid model corresponds to the presence of a cross-over transition. In Fig. 3, we have separately shown the variation of square of speed of sound i.e., $c_{s}^{2}$ for HG and  also for QGP. We have also shown the difference $(\Delta c_{s}^{2})=(c_{s}^{2})_{QGP}- (c_{s}^{2})_{HG}$ and demonstrated its variations with respect to the coordinates of the phase transition points lying at the deconfining phase boundary. We again find that $\Delta c_{s}^{2}\approx 0.0$ at the CP. Thus this results lends further support to our previous finding regarding the change of the order of the transition at the critical point.

In Fig. 4 (A), we plot the variation of $\eta/s$ of gluon plasma with respect to $T/T_{c}$, where $T_{c}$ is the critical temperature. We compare our model result with the results obtained by Khvorostukin et. al. [35] and Bluhm et. al. [36]. We have also shown a comparison with the lattice gauge calculations [37, 38]. We observe that the results obtained in our calculation agrees well with the lattice data even at large temperatures. However, the error in lattice simulation result as given in Fig. 4 (A) is quiet large. Our result lies in close aggreement with the result obtained by Khvorostukin et. al [35]. This result naturally gives us additional confidence to calculate $\eta/s$ of QGP by using QPM. In Fig. 4 (B), we plot the variation of shear viscosity to entropy density ratio for HG and for QGP, respectively with respect to temperature at $\mu_{B}=300$ MeV (dashed-line) and  at critical potential $\mu_{c}=155$ MeV (solid-line). At $\mu_{B}=300$ MeV, we observe a discontinuity in $\eta/s$ at the joint of both the curves for phases i.e., HG and QGP, at first order phase transition point. Further we observe an upward jump in $\eta/s$ going from low temperature HG phase to high temperature QGP phase which is in accordance with the result obtained by Sasaki and Redlich [13]. At critical chemical potential  $\mu_{c}=155$ MeV and temperature $T_{c}=166$ MeV [9] we get a cusp like behaviour in $\eta/s$ while going from HG to QGP phase as shown in Ref. [13, 39].
 
 Thus above results give a firm indication that the order of phase transition changes at CP for the deconfinement phase transition. It should be added here that many authors in the past have used two separate equations of state for QGP and HG and obtained a tentative explanation for an analytic and smooth cross-over and CP in their models [10-11]. Our model presents a similar picture. The physical mechanism involved in this calculation is intuitively analogous to the percolation model where a first order phase transition results due to 'jamming' of baryons which thus restricts the mobility of baryons [40-41]. However, in the percolation model we do not have any comparison to what one gets in the QGP picture. Here we explicitly and separately consider both the phases, i.e., HG as well as QGP and hence it gives a clear understanding how a first-order deconfining phase transition can be constructed in nature. At low baryon density, overlapping mesons fuse into each other and form a large bag or cluster, whereas at high baryon density, hard-core repulsion among baryons, restricts the mobility of baryons. Consequently we consider two distinct limiting regimes of HG, one is a meson-dominant regime and the other is a baryon dominant region and the point joining the two is CP in our models.

In conclusion, searching for the precise location and the nature of the critical point (CP) on the QCD phase diagram still poses a challenging problem. Although various calculations have predicted its existence but the quantitative predictions regarding its location wildly differ. Experiments face an uphill task in probing the CP in QCD phase diagram because a clarity in theoretical prediction is missing. Moreover, many unstudied problems such as short lifetime and the reduced volume of the QGP formed at colliders also affect the location of CP and its verification [42]. In these circumstances, we believe that our results will throw light on the nature of CP existing in deconfining phase transition.\\

\noindent
{\bf Acknowledgments}\\

 PKS is grateful to the University Grants Commission (UGC), New Delhi for providing a research grant. CPS acknowledges the financial support through a project sanctioned by Department of Science and Technology, Government of India, New Delhi.

\pagebreak

\end{document}